\documentclass[conference]{IEEEtran}
\IEEEoverridecommandlockouts
\usepackage{cite}
\usepackage{amsmath,amssymb,amsfonts}
\usepackage{tabularx}
\usepackage{algorithm}
\usepackage{algpseudocode}
\usepackage{rotating} % 在导言区添加
\usepackage{graphicx}
\usepackage{textcomp}
\usepackage{xcolor}
\usepackage{booktabs}
\usepackage{subfig}
\usepackage{dsfont}
\usepackage{array,color}
\usepackage{bm}
\usepackage{float}
\usepackage{multirow}
\usepackage{multirow}
\usepackage{makecell}
\usepackage{tikz}
\usetikzlibrary{positioning} %为了实现相对位置的设定
\usepackage{amsfonts}
\usepackage{tgtermes}
\usepackage{tgheros}
\usepackage{cite,graphicx,amsmath,amsthm}
\usepackage{amsmath} 
\usepackage{url}
\begin{document}
\title{LLM Agent for Hyper-Parameter Optimization}
\author{
Wanzhe Wang, Jianqiu Peng,  Menghao Hu, Weihuang Zhong, Tong Zhang, \\
Shuai Wang, Yixin Zhang, Mingjie Shao, and Wanli Ni

\thanks{Wanzhe Wang, Jianqiu Peng,  Weihuang Zhong are with the Harbin Institute of Technology (Shenzhen), Shenzhen, China (email: wangwz724@163.com, \{13400053106,15986817267\}@163.com).}
\thanks{Menghao Hu is with the School of Information Science and Engineering, Southeast University, China (email: 220240758@seu.edu.cn).}
\thanks{Tong Zhang is with the Guangdong Provincial Key Laboratory of Aerospace
Communication and Networking Technology, Harbin Institute of Technology,
Shenzhen 518055, China, and also with the National Mobile Communications Research Laboratory, Southeast University, China (e-mail: tongzhang@hit.edu.cn).}
\thanks{Shuai Wang is with the Shenzhen Institutes of Advanced Technology, Chinese Academy of Sciences, Shenzhen, China (email: s.wang@siat.ac.cn).}
\thanks{Yixin Zhang is with the Beijing University of Posts and Telecommunications, Beijing, China (email: yixin.zhang@bupt.edu.cn).}
\thanks{Mingjie Shao is with the State Key Laboratory of Mathematical Sciences, AMSS, Chinese Academy of Sciences, Beijing, China (email: mingjieshao@amss.ac.cn).}
\thanks{Wanli Ni is with the Department of Electronic Engineering, Tsinghua University, Beijing 100084, China (e-mail: niwanli@tsinghua.edu.cn).}
\thanks{Corresponding author: T. Zhang}
}
\maketitle

\begin{abstract}
Hyper-parameters are essential and critical for the performance of communication algorithms.
However, current hyper-parameters optimization approaches for Warm-Start Particles Swarm Optimization with Crossover and Mutation (WS-PSO-CM) algorithm, designed for radio map-enabled unmanned aerial vehicle (UAV) trajectory and communication, are primarily heuristic-based, exhibiting low levels of automation and improvable performance. In this paper, we design an Large Language Model (LLM) agent for automatic hyper-parameters-tuning, where an iterative framework and Model Context Protocol (MCP) are applied. In particular, the LLM agent is first set up via a profile, which specifies the boundary of hyper-parameters, task objective, terminal condition, conservative or aggressive strategy of optimizing hyper-parameters, and LLM configurations. Then, the LLM agent iteratively invokes WS-PSO-CM algorithm for exploration. Finally, the LLM agent exits the loop based on the terminal condition and returns an optimized set of hyperparameters. Our experiment results show that the minimal sum-rate achieved by hyper-parameters generated via our LLM agent is significantly higher than those by both human heuristics and random generation methods. This indicates that an LLM agent with PSO  and WS-PSO-CM algorithm knowledge is useful in seeking high-performance hyper-parameters.
\end{abstract}


\section{Introduction}

The rapid advancement of unmanned aerial vehicles (UAV) has revolutionized wireless communication systems, offering flexible deployment and enhanced coverage in dynamic environments \cite{Zhaohui,Menghao1}. A critical enabler for UAV-assisted communication is the radio map, which provides essential spatial channel information to optimize trajectory planning and transmission strategies. Among the optimization algorithms, Warm-Start Particle Swarm Optimization with Crossover and Mutation (WS-PSO-CM) \cite{hu2025radiomapenabled3dtrajectory} has emerged as a promising approach due to its adaptability and efficiency in solving complex UAV trajectory and communication problems. However, the performance of WS-PSO-CM heavily relies on the selection of hyper-parameters, which remains a significant challenge.

Radio map aims to reflecting the spatial distribution characteristics of wireless channel, which can construct models of radio signal distribution in the environment by collecting and analyzing information about signal strength, delay, and channel state. Yang \textit{et al.}~\cite{10272348} proposed a deep learning-based radio map that directly generates beamforming vectors from location information using a task-oriented neural network. In \cite{9707858}, Dong \textit{et al.} presented radio maps in UAV anti-interference communications. In \cite{9269485}, Zhang \textit{et al.} proposed the use of radio maps for cellular network connectivity in UAV three-dimensional path planning, aiming to address the issues of communication quality assurance and path optimization during UAV flight. In \cite{11008499}, Yang \textit{et al.} proposed an end-to-end method integrating radio map-based beamforming with reduced pilots. In \cite{9513834}, Yuan \textit{et al.} introduced radio maps into the design of multi-UAV relay networks and employed a joint optimization approach using the particle swarm optimization (PSO) algorithm to overcome the limitations of traditional models in complex geographical environments. In \cite{1}, Dong \textit{et al.} addressed the problem of dynamic multi-UAV target tracking assisted by radio maps by proposing methods based on grid and particle filtering techniques. In \cite{10032196}, the PSO is integrated with the genetic algorithm (GA) for 3D UAV trajectory optimization, considering obstacles. However, when using a dynamic radio map, specifically a large path-loss (PL) database instead of statistical channel models, the computational time required for PSO combined with GA becomes excessively high, leading to suboptimal performance when time constraints are imposed. In \cite{hu2025radiomapenabled3dtrajectory}, Hu \textit{et al.} introduced WS-PSO-CM algorithms, which uses statistical channel models for convex optimization to generate a high-quality initial particle swarm for warm start (WS). Compared to the baseline algorithm PSO-CM, it significantly improves performance within the same number of iterations. 

Currently, hyper-parameter tuning for WS-PSO-CM is predominantly heuristic-based, requiring extensive domain expertise and manual intervention. Such methods not only lack automation but also often yield suboptimal configurations, limiting the algorithm’s potential in real-world applications.  

Large Language Model (LLM), as a significant breakthrough in artificial intelligence (AI), demonstrates powerful capabilities in knowledge representation and task inference, enabling novel integrative applications across diverse domains~\cite{qiu2024llm,romera2024mathematical,ren2024industrial}. In \cite{qiu2024llm}, Qiu \textit{et al.} envisioned that the LLM agent system enhanced the accuracy of medical diagnostics and in \cite{romera2024mathematical} Romera-Paredes \textit{et al.} proposed that LLM enabled the automated exploration of complex mathematical problems and algorithm optimization tasks. In \cite{ren2024industrial}, Ren \textit{et al.} introduced the application of LLM in industrial internet of things. The authors of \cite{10763424} proposed low-rank adaptation-based federated fine-tuning for large models in wireless networks and derived an optimality gap to capture the impact of rank selection and gradient aggregation distortion. Furthermore, the authors of \cite{ni2025federated} presented multi-agent collaborative reasoning frameworks to enhance efficiency of large AI models. In \cite{qiu2024large}, Qiu \textit{et al.} employed LLM-enabled optimization for wireless network planning, demonstrating that large language models can significantly enhance the optimization process in wireless network design. The advanced capabilities of LLM agents enables the development of more automated and intelligent systems capable of generating precise radio maps and planning wireless networks with minimal human intervention. In \cite{10758700}, Chatzistefanidis  \textit{et al.} proposed a collaborative framework utilizing LLMs to achieve greater abstraction and automation in network planning, thereby substantially streamlining the deployment process. Also, in \cite{10757657}, Sevim \textit{et al.} adopted an alternative approach by developing a reinforcement learning-based system enhanced with LLM capabilities for wireless network deployment. In \cite{NL}, Quan \textit{et al.} utilized the capability of LLM for radio map generation and wireless network planning.

In this paper, we investigate an automatic hyper-parameters-tuning LLM agent for WS-PSO-CM algorithm, specifically tailored for radio map-enabled UAV trajectory and communication. In particular, to build the LLM agent, we apply model context protocol (MCP), which is an open protocol that standardizes how applications provide context to LLMs. We employ an iterative optimization framework for our proposed LLM agent with PSO and WS-PSO-CM algorithm knowledge, where the agent engages in sequential exploration of hyper-parameter space through dynamic interactions with WS-PSO-CM algorithm. After a few iterations, our proposed LLM agent outputs a set of hyper-parameters. Experiment results show that the obtained hyper-parameters can achieve $54.33\%$ and $72.61\%$ gains over that obtained by human heuristics in \cite{hu2025radiomapenabled3dtrajectory} and an uniform distribution, respectively.

\section{Optimization of 3D UAV Trajectory and Communication: Introduction of WS-PSO-CM Algorithm}

\subsection{System Model and Problem Formulation}
In this section, we considered the WS-PSO-CM algorithm for joint optimization of UAV trajectories and communications for low-altitude air-ground cooperation \cite{hu2025radiomapenabled3dtrajectory}. 
The problem WS-PSO-CM algorithm is aroused by an air-ground cooperation system and $N$ unmanned ground vehicles (UGVs) communicate with $M$ UAVs. We represent the 3D positions of UAVs and UGVs at each time slot $t$ as ${\bm{\ell}}^{\text{\text{UAV}}}_{m}[t]=(\bm{\alpha}_{m}[t],H_{m}[t])=(x_{m}^{\alpha}[t], y_{m}^{\alpha}[t], H_{m}[t])$ and ${\bm{\ell}}^{\text{UGV}}_{n}[t]=(\bm{\beta}_{n}[t],0)=(x_{n}^{\beta}[t], y_{n}^{\beta}[t], 0)$, respectively, where $m \in \mathcal{M}=\{1,2,...,M\}$, $n \in \mathcal{N}=\{1,2,...,N\}$. To prevent collisions, the distance between UAVs must exceed a specified minimum threshold, i.e.,
\begin{equation}
\label{anti-collision}
\lVert{\bm{\ell}}^{\text{UAV}}_{m}[t] - {\bm{\ell}}^{\text{UAV}}_{m'}[t] \rVert_2 \geq d_{\min}, \,\,\, m \neq m',  \,\forall m,m',t.
\end{equation}		
Similar to \cite{hu2025radiomapenabled3dtrajectory}, the author introduce a binary scheduling variable $a_{m,n}[t] \in \{0,1\}$, where $a_{m,n}[t]=1$, if UGV $n$ communicates with UAV $m$ at time slot $t$, and otherwise $a_{m,n}[t]=0$. Furthermore, at each time slot, each UAV communicates with only one UGV, i.e.,
\begin{equation}
\label{TDMA1}
\sum_{n=1}^{N}a_{m,n}\left[t\right]\leq1, ~\sum_{m=1}^{M}a_{m,n}\left[t\right]\leq 1,\,\,\, \forall m,n,t. 
\end{equation}
	 
The radio map, i.e., PL map, for UGV $n$ acting as a transmitter in time slot $t$ is denoted by $\textbf{H}_{n}[t] \in \mathbb{R}^{X \times Y \times Z}$, where the 3D physical space is divided into cubic cells $X \times Y \times Z$, each with a volume of $\delta^3$. The PL values in map $ \textbf{H}_{n}[t]$ at the 3D coordinate index $ (x_{m},y_{m},z_{m})$ is denoted by $h_{n,t}(\omega_m)$, where the index $\omega_m = (x_{m},y_{m},z_{m})$. This index can be mapped to the actual position of a UAV using the following relations: $ x_{m}=\left\lfloor  (x_{m}^{\alpha}[t]-X_{\min})/\delta \right\rfloor +1$, $ y_{m} =\left\lfloor  (y_{m}^{\alpha}[t]-Y_{\min})/\delta \right\rfloor +1$, and $z_{m} =\left\lfloor (H_{m}[t]-H_{\min})/\delta \right\rfloor+1$, where $X_{\min}$, $Y_{\min}$ and $H_{\min}$ serve as anchor points,  and $H_{\min}$ also represents the minimum flight altitude of the UAVs. The rate between UGV $n$ and UAV $m$ at time slot $t$ is calculated as $ R_{m,n}\left[t\right] = \log_2\left(1+a_{m,n}[t] \text{SINR}_{m,n}[t]\right)$, where { $\text{SINR}$ $=  \frac{h_{n,t}(\omega_m)P_{n}[t]}{\sum_{p\neq n}^{N}(\sum_{q=1}^{M}a_{pq}[t])h_{p,t}(\omega_m)P_{p}[t]+N_0}$,} with $N_{0}$ denoting the additive white Gaussian noise (AWGN) power.

In order to maximize the minimal sum-rate of UGVs subject to scheduling scheduling, power, quality of service and UAV flight constraints, we optimize the UGV-UAV scheduling set $\mathcal{A}=\{a_{m,n}\left[t\right]\}$, UAV 3D trajectory set $\mathcal{Q}=\{\bm{\alpha}_{m}[t]\}$ and UGV transmit power set $\mathcal{P}=\{P_{n}\left[t\right]\}$. Mathematically, this problem can be formulated as the following mixed-integer programming problem:
\begin{subequations}		
			\begin{eqnarray}
		\!\!\!\!\!	(\text{P0})	 \,\,\,				 \max_{\mathcal{A},\mathcal{Q},\mathcal{P}}~&&  \!\!\!\!\!\!\!\!\! \min \frac{1}{T}\sum_{t=1}^{T}\sum_{m=1}^{M} R_{m,n}[t]   \\
				\textrm{s.t.}~~ 
								\nonumber&&\!\!\!\!\!\!\!\!\!{\eqref{anti-collision}}, {\eqref{TDMA1}},\\
				\label{qos}&&\!\!\!\!\!\!\!\!\! R_{m,n}[t]\geq a_{m,n}[t]R_{\min},  \forall m,n,t,\\
				%\label{lamda} && \sum_{t=1}^{T}\sum_{m=1}^{M} R_{m,n}[t]\geq \mu, \, \forall n,\\
				%	\label{anti-collision}&\lVert\bm{\alpha}_{m}[t]-\bm{\alpha}_{q}[t] \rVert^{2} \geq d^{2}_{\min}, m \neq q,  \forall t \\
				\label{max-speed}&&\!\!\!\!\!\!\!\!\! \lVert{\bm{\ell}}^{\text{UAV}}_{m}[t+1]-{\bm{\ell}}^{\text{UAV}}_{m}[t] \rVert_2 \leq V_{\max}\tau,t <T, \\
			  	\label{HeightC}&&\!\!\!\!\!\!\!\!\! H_{\min} \leq H_{m}[t] \leq H_{\max}, \forall m, t,\\
			  	\label{angle}&&\!\!\!\!\!\!\!\!\!  {\theta_{m}[t] \leq \theta_{\max}, \forall m, t,}\\
			  	\label{building}&&\!\!\!\!\!\!\!\!\! { \bm{\ell}^{\text{UAV}}_{m}[t] \notin \mathcal{B}, \forall m, t,}\\
				\label{power}&&\!\!\!\!\!\!\!\!\! 0 \leq P_{n}[t] \leq P_{\max}, \forall n,t,\\
				\label{0-1}	&&\!\!\!\!\!\!\!\!\! a_{m,n}[t] \in\left\{0,1\right\}, \forall m,n,t,	 
			\end{eqnarray}			\label{P0}
		\end{subequations}
\!\!\!where constraint {\eqref{qos}} enforces the quality of service requirement for each communication link, where the minimum acceptable data rate is specified by $R_{\min}$. The UAVs’ velocity is limited to a maximum of $V_{\max}$ by constraint {\eqref{max-speed}}, with $\tau$ denoting the length of each time slot. The UAVs’ flight altitude is restricted to remain within the bounds of $H_{\max}$ and $H_{\min}$ by constraint {\eqref{HeightC}}. Constraint {\eqref{angle}} imposes a limit on the UAVs’ turning angle, which is not allowed to exceed the maximum allowable value $\theta_{\max}$; here, $\theta_{m}[t]$ is defined as the angle between ${\bm{\ell}}^{\text{UAV}}{m}[t]-{\bm{\ell}}^{\text{UAV}}{m}[t-1]$ and ${\bm{\ell}}^{\text{UAV}}{m}[t+1]-{\bm{\ell}}^{\text{UAV}}{m}[t]$. Collision avoidance between UAVs and buildings is guaranteed by constraint {\eqref{building}}, where $\mathcal{B}$ denotes the three-dimensional space occupied by buildings. Finally, constraint {\eqref{power}} ensures that the transmit power of UGVs does not exceed $P_{\max}$.

The authors of \cite{hu2025radiomapenabled3dtrajectory} proposes a general WS-PSO-CM algorithm for radio maps, which will be validated through simulations using the established radio map database.

\subsection{WS-PSO-CM Algorithm}

The standard PSO algorithm performs global optimization by simulating collective biological behavior, where particles (representing candidate solutions) move through the solution space with dynamically updated velocities based on individual and social learning experiences, gradually converging toward the global optimum.  WS-PSO-CM algorithm enhances standard PSO through two key modifications: a warm-start initialization strategy that dramatically reduces convergence time, and the integration of genetic algorithm operators (crossover and mutation) to sustain solution diversity throughout the optimization process.

WS-PSO-CM algorithm is used for optimizing the UAV trajectory and communication, considering several objectives like minimizing path loss, avoiding collisions, adhering to speed limits, and maximizing the minimal sum-rate of communication links between UAVs and UGVs.

\subsubsection{Initialization} A population of particles is initialized. These particles represent potential solutions.Also, we define that $P_\text{iter}$ and $P_\text{num}$ are the number of iteration and number of particles in the proposed WS-PSO-CM. 

\subsubsection{Main Optimization Loop} Particles update their velocity by considering their current velocity, the global best solution, and their local best solution. Based on this, they adjust their position accordingly. The strategy for updating both position and velocity and the fitness function is as follows:
	\begin{subequations}
		\begin{align}
			&\mathcal{V}_{k}=\omega \times\mathcal{V}_{k} + c_{1}\times\textsc{rand}_{1}\times\left(pBest_{k}-\mathcal{V}_{k}\right)\label{updateV}\\
			&\phantom{\mathcal{V}_{k}=}+c_{2}\times\textsc{rand}_{2}\times\left(gBest-\mathcal{V}_{k}\right),\notag\\
			&\mathcal{Q}_{k}=\mathcal{Q}_{k}+\mathcal{V}_{k}\label{updateQ},\\
            &F(\mathcal{Q}_{k}) = {k}_1 T(\mathcal{Q}_{k})+{k}_2 S(\mathcal{Q}_{k})+ {k}_3 A(\mathcal{Q}_{k})+{k}_4 C(\mathcal{Q}_{k}),
		\end{align}
	\end{subequations}	
where $ \omega$, $ c_{1}$ and $ c_{2}$ represent the inertia weight, cognitive coefficient, and social coefficient, respectively, which influence the update of the particles' position and velocity. $pBest_k$ and $ gBest$ are the $k${th} particle's position and the best particle's position at current iteration. $ \textsc{rand}_{1}$ and $ \textsc{rand}_{2}$ are two random variables ranging from $[0,1]$ with uniform distribution. Also, where $k_{1}$, $k_{2}$, $k_{3}$, $k_{4}$ are the weight coefficients. $T(\mathcal{Q}_{k})=\min\sum_{t=1}^{T}\sum_{m=1}^{M} R_{m,n}[t]$ is denoted as the min sum-rate. The sum of the normalized velocities exceeding the maximum speed limit $V_{\max}$ is denoted by $S(\mathcal{Q}_{k})$, $S(\mathcal{Q}_{k}) = \sum_{m=1}^{M}\sum_{t=2}^{T} \max (0, \frac{v_{m}[t] - V_{{\max}}}{V_{{\max}}} )$
	with {$ v_{m}[t] = \frac{\lvert {\bm{\ell}}^{\text{UAV}}_{m}[t]-{\bm{\ell}}^{\text{UAV}}_{m}[t-1]\rvert}{\tau}$};
  $ A(\mathcal{Q}_{k})$ is denoted as the sum of the normalized turning angle exceeding the maximum {limit}, i.e., $A(\mathcal{Q}_{k}) = \sum_{m=1}^{M}\sum_{t=2}^{T-1} \max (0, \frac{\theta_{m}[t] - \theta_{{\max}}}{\theta_{{\max}}} )$.  
  $ C(\mathcal{Q}_{k})$ is denoted as the sum of the amount of height of path points located within $\mathcal{C}_{b}$ exceeding the height of the building, i.e. $C(\mathcal{Q}_{k}) = \sum_{m=1}^{M}\sum_{t=1}^{T-1}\max\left(0, \Delta_{m}[t]\right)$,

\subsubsection{Crossover and Mutation} Crossover and mutation operations are applied periodically to the particles to introduce genetic diversity and explore new regions of the solution space.
\subsubsection{Stopping Condition}The algorithm stops after a specified number of iterations and the best solution is returned.

\begin{algorithm}[t]
\caption{Pseudo Code of WS-PSO-CM Algorithm}
\begin{algorithmic}[1]
\State \textbf{Initialize:} Warm start by convex optimization (see \cite{hu2025radiomapenabled3dtrajectory}), set iter = 1, and $P_\text{num}$, $\omega$, $c_{1}$, $c_{2}$, $k_{1}$, $k_{2}$, $k_{3}$, $k_{4}$;
\State\textbf{Repeat}:
    \State \quad Update particles' velocities according to {\eqref{updateV}};
    \State \quad Update particles' positions according to {\eqref{updateQ}};
    \State \quad Update the $\rho Best_k$;
    \State \quad Select $gBest$ from $\rho Best_k$;
    \State \quad Particles cross and mutate according to Cross and Mutation strategy;
    \State \quad Set iter $\gets$ iter + 1;
\State\textbf{Until:}iter = $P_{iter}$;
\State \textbf{Output:} $gBest$;
\vspace{-0.15cm}
\end{algorithmic}
\end{algorithm}
Here, $P_\text{num}$, $\omega$, $c_{1}$, $c_{2}$, $k_{1}$, $k_{2}$, $k_{3}$, $k_{4}$ are hyper-parameters within the algorithm. After obtaining an optimized result from a given initial value, the analysis of the optimization outcome is performed using LLM to adjust the hyper-parameters, which in turn allows for further refinement of the previous results.

\begin{figure*}[t]
  \centering
  \includegraphics[width=0.75\textwidth]{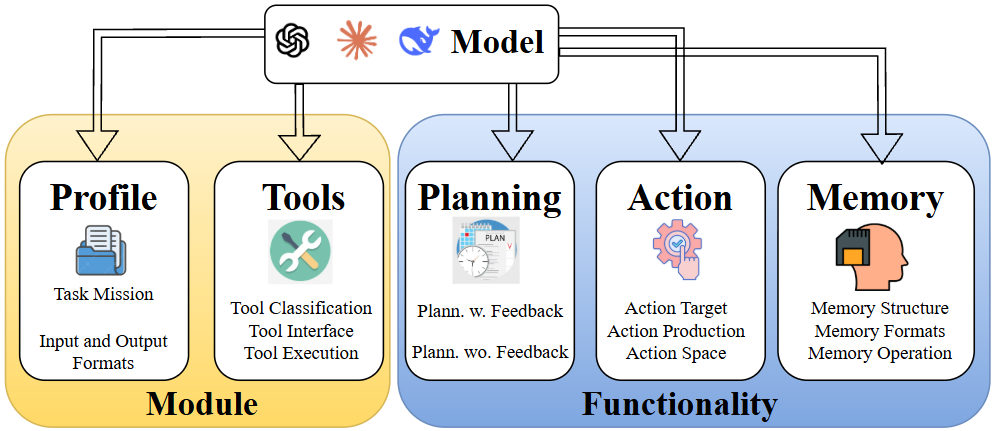}
  \caption{LLM agent modules and functionalities.}
  \label{LLM}
\end{figure*}

\section{MCP and Proposed LLM Agent}

\subsection{A Brief Introduction to LLM Agent}

Besides LLM model, the LLM agent framework through two modules (Profile, Tools) and three functions (Planning, Action, Memory), presented in Fig.~\ref{LLM}, which are elaborated below.

\subsubsection{Profile}
The Profile module serves as the foundational context for the agent's understanding of the user. It encompasses profile context, which includes demographic, personality, and social information about the user, providing a personalized basis for task execution. The generation strategy within this module determines the approach by which the agent generates responses, utilizing methods such as handcrafted techniques, LLM generation approaches, or dataset alignment methods to tailor output according to specific user needs.

\subsubsection{Tools}
The Tools module provides external application interface (API) integration for enhanced agent capabilities. It is the agent’s interface with external systems, enabling it to perform operations beyond its native LLM capabilities. It acts as a secure, structured gateway for integrating APIs, databases, and computational services while maintaining control over execution.

\subsubsection{Planning}
In Planning functionality, the agent formulates strategies for task execution. This process can occur without Feedback, where the agent employs single-path reasoning, multi-path reasoning, or external planning to generate an initial action plan. Alternatively, planning with feedback integrates feedback from the environment, human input, or the model itself, allowing the agent to refine its plan and adjust its approach dynamically in response to changing conditions.

\subsubsection{Action}
The Action functionality involves the execution of the agent’s formulated plans. The action target determines the target of the action, which can range from task completion and communication to exploration. Action production is driven by agent memory recall, where it generates concrete actions based on prior knowledge. The action space defines the scope of possible actions, incorporating tools and the agent’s own knowledge to facilitate the execution of the plan.

\subsubsection{Memory}
The Memory functionality is central to the agent’s ability to store and recall relevant information over time. It is structured around a memory structure, which can either be unified or hybrid, depending on the system's requirements for knowledge management. The memory formats include various types of stored information, such as languages, databases, embeddings, and lists, each serving a distinct purpose for task execution. The memory operation allows the agent to interact with stored knowledge through processes like memory reading, writing, and reflection, enabling dynamic updating and retrieval of information during task completion.

\subsection{MCP and Proposed LLM Agent}

\subsubsection{MCP Architecture}
MCP introduces a standardized communication framework that simplifies the development of LLM-based agents. By providing a unified protocol, it enables seamless integration of language models into agent systems. The core idea of MCP is to abstract the communication between models and external systems into a client-server architecture, enabling dynamic context passing and flexible tool invocation through standardized interfaces (e.g., JSON-RPC-based communication). This architecture allows a MCP client to establish connections with multiple servers, thereby facilitating flexible context transfer and functional extensibility.

\begin{figure}[t]
  \centering
  \includegraphics[width=0.5\textwidth]{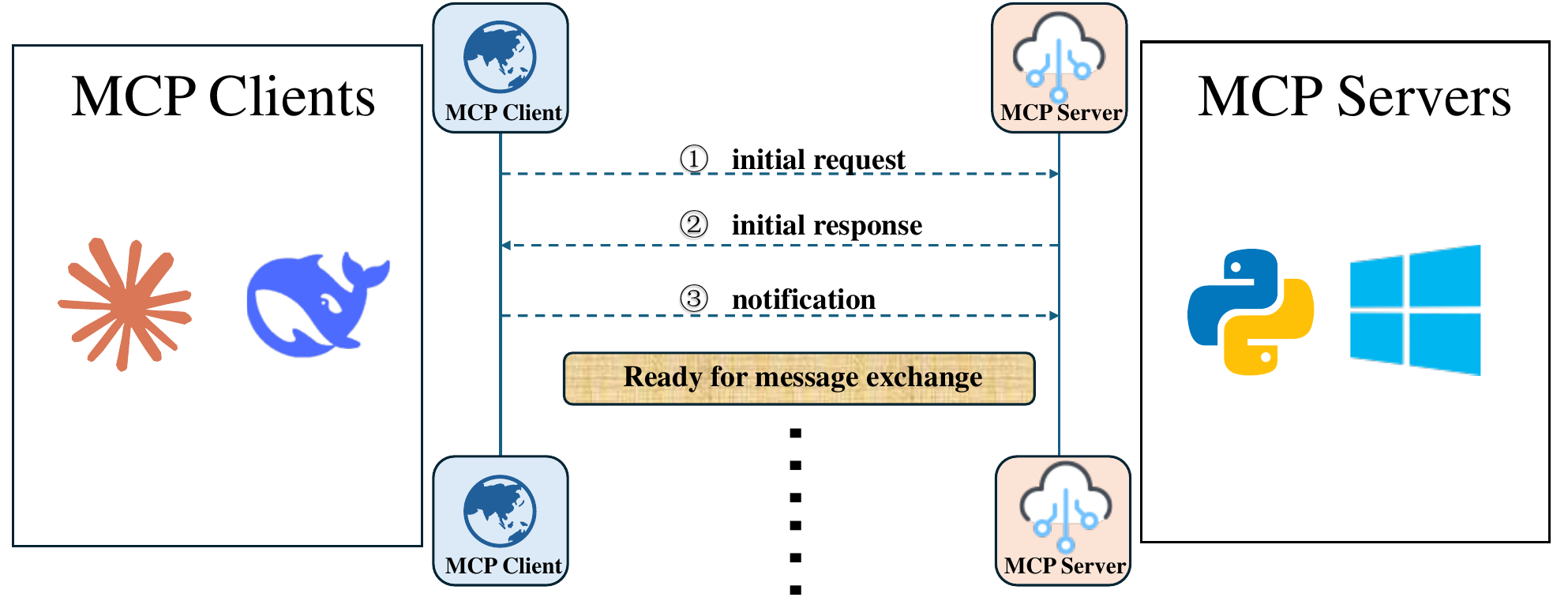}
  \caption{Illustration of MCP.}
  \label{transport layer}
\end{figure}

Fig.~\ref{transport layer} shows the MCP architecture, responsible for enabling secure bidirectional communication between MCP client and MCP server, which is developed by Anthropic. MCP server functions as a centralized tool runtime controller, responsible for authenticating, scheduling, and monitoring registered tools (e.g., APIs), whereas MCP client acts as a tool invocation proxy that initiates requests and formats tool inputs/outputs according to protocol specifications. While MCP server maintains tool metadata repositories and enforces execution policies (e.g., rate-limiting, dependency resolution), MCP client focuses on context-aware tool selection and lightweight result post-processing, delegating heavy tool operations to MCP server.

%The specific steps of the communication process: 

%\subsubsection{Features}
%The MCP protocol is an efficient communication protocol designed for large-scale machine learning models and distributed training environments, aimed at addressing the challenges of data transfer and model parameter synchronization between multiple computing nodes during training. As the scale of deep learning models continues to expand, especially with large pre-trained models like GPT and BERT, traditional single-device training methods can no longer meet computational demands. Therefore, the MCP protocol was introduced to solve the efficiency and stability of multi-node collaborative work.

Also, MCP optimizes communication paths and transmission formats, reducing communication latency in distributed environments, especially when handling large-scale gradient updates and model parameters. It supports both synchronous and asynchronous training modes, allowing flexible adjustments based on training needs, thereby improving training efficiency while ensuring model accuracy. Additionally, the MCP has a fault tolerance mechanism, ensuring stable operation of the training process even in the event of network fluctuations or hardware failures, with automatic recovery in case of abnormalities, preventing the entire training process from being interrupted due to a single point of failure.

Security and privacy protection are a major highlight of the MCP protocol, especially in scenarios such as federated learning. When multiple institutions or entities collaborate on training, the protocol ensures that data is not directly shared while employing encryption methods to protect data privacy during communication. In response to the diversity of modern computing resources, the MCP protocol also offers intelligent resource scheduling and load balancing mechanisms, enabling different computing nodes to work efficiently together, maximizing the use of computational resources and avoiding resource wastage and performance bottlenecks.

The interaction process between the MCP client and the MCP server is structured and systematic. Initially, the MCP client is configured to establish a connection with a designated MCP server. Upon receiving a user's prompt, the MCP client formulates a structured prompt that incorporates both the user's intent and the tools made available by the MCP server. This composite input is then submitted to the LLM for interpretation and task planning. The LLM interprets the prompt and generates corresponding invocation instructions. These instructions are then transmitted from the client to the MCP server. Upon receipt, the MCP server parses the request, executes the appropriate operations.

Overall, the MCP protocol not only provides an efficient communication means for the training of large-scale deep learning models but also promotes the application of multi-party collaboration and privacy protection technologies, laying a solid foundation for the future development of AI.

\subsubsection{Proposed LLM Agent}
\begin{figure}[t]
  \centering
  \includegraphics[width=0.48\textwidth]{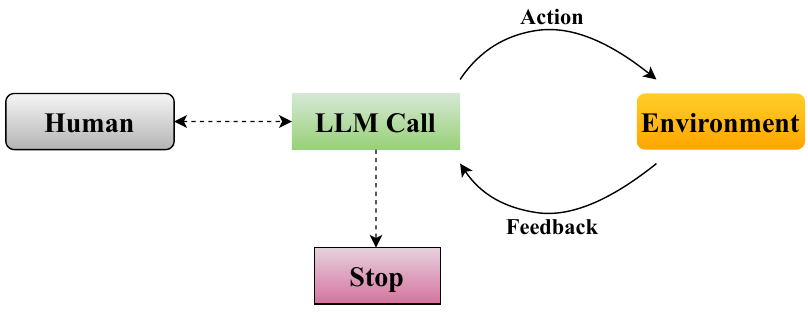}
  \caption{The framework of proposed LLM agent.}
  \label{LLM_0}
\end{figure}

As shown in Fig.~\ref{LLM_0}, we propose an interactive LLM agent that orchestrates collaboration between the model, humans, and the environment, namely WS-PSO-CM algorithm. 
In the profile, we define the boundary of hyper-parameters, task objective, terminal condition, conservative or aggressive strategy of optimizing hyper-parameters, and LLM configurations. Tools include two MCP servers. The first server specializes in hyper-parameter transmission, delivering optimized parameters to numerical computation software for WS-PSO-CM operations. The second server focuses on hyper-parameter management and historical record maintenance. 

Once the proposed LLM agent runs, the LLM generates hyper-parameters based on the profile. The algorithm’s optimization results are fed back to the LLM for iterative refinement. The LLM agent terminates the loop according to prescribed conditions and returns a set of hyper-parameters.

%\section{Programming Development}
\section{Experiments}

In this experiment, to build LLM agent, we choose deepseek R1 as MCP client. Our simulation scenario covers the teaching building area and hall of HITSZ, with dimensions of $240 \times 400 \times 60$ m$^3$. A corresponding digital twin is then constructed in mathematical software for numerical calculations using PL map data. Our simulation parameters are the same as that in \cite{hu2025radiomapenabled3dtrajectory}, except the number of UGVs is $8$, the number of UAVs is $4$, the speed of UGV 1, 2, 3, 4, 5, 6, 7, 8 are 18.0km/h, 18.0km/h, 13.8km/h, 13.8km/h, 12.5km/h, 17.0km/h, 13.0km/h, 19.0km/h, respectively. In this experiment, the hyper-parameters awaiting tuning are as follows: $P_\text{num}$, $\omega$, $c_{1}$, $c_{2}$, $k_{1}$, $k_{2}$, $k_{3}$, $k_{4}$. We adopt the following two baselines for comparison:
\begin{itemize}
    \item {\underline{Baseline 1} (Human Heuristics in \cite{hu2025radiomapenabled3dtrajectory})}: These hyper-parameters are used in \cite{hu2025radiomapenabled3dtrajectory} based on human heuristics, where $P_\text{num}=100$, $\omega=0.5$, $c_{1}=2$, $c_{2}=2$, $k_{1}=2$, $k_{2}=0.5$, $k_{3}=5$, $k_{4}=5$.
    \item {\underline{Baseline 2} (Random Generation via Uniform Distribution)}: Generate hyper-parameters via uniform distribution, where $P_\text{num}=58$, $\omega=0.8765$, $c_{1}=5.4321$, $c_{2}=9.8765$, $k_{1}=3.7284$, $k_{2}=8.1235$, $k_{3}=1.9823$, $k_{4}=6.5432$.
\end{itemize}

Since we only consider iterations within a few rounds, we do not include Grid Search and Bayesian Optimization in the comparison.
 
\begin{figure}[t]
    \centering
    \includegraphics[width=0.7\linewidth]{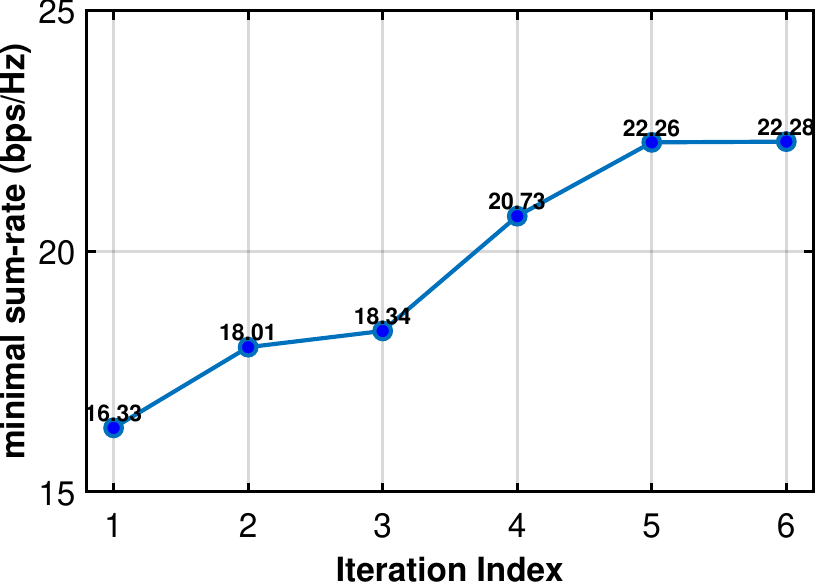}
    \caption{Minimal sum-rate over iterations by LLM agent.}
    \label{fig1}
\end{figure}

In Fig. \ref{fig1}, it shows the minimal sum-rate improvement over iterations by LLM agent. In each iteration, the recommended hyper-parameters are given in TABLE I.  It can be seen that by reading the WS-PSO-CM algorithm and PSO background, LLM agent yields satisfactory hyper-parameters within $6$ iterations. 
Note that, the human heuristics hyper-parameters in \cite{hu2025radiomapenabled3dtrajectory} are with $100$ particles. Meanwhile, TABLE I demonstrates that the recommended number of particles does not exceed $50$. Next, by observing the output of WS-PSO-CM algorithm, the LLM agent keeps refining the hyper-parameters. It stops until a minimal sum-rate of $22.28$ bps/Hz is achieved. This shows that with PSO background and WS-PSO-CM algorithm, the LLM agent explores the landscape of the optimization problem, and gives high-performance hyper-parameters eventually.

In Fig. \ref{fig1}, it shows that our proposed LLM agent achieves $54.33\%$ and $72.61\%$ gains over that by hyper-parameters via human heuristics in \cite{hu2025radiomapenabled3dtrajectory} and an uniform distribution, respectively. This demonstrates that the LLM agent equipped with PSO and WS-PSO-CM algorithm knowledge possesses strong capabilities in finding high-performance hyper-parameters, which are crucial for optimizing WS-PSO-CM algorithm performance. With the hyper-parameters by LLM agent, the optimized UAV 3D trajectory is visualized in Fig.  \ref{Topview}.

\begin{table}[t]
\centering
\caption{Hyper-Parameters during Iterations}
\label{experiments}
%\small  % 减小字体大小
\begin{tabular}{ccccccccc}
\toprule
Index & $P_\text{num}$ & $k_1$ & $k_2$ & $k_3$ & $k_4$ & $w$ & $c_1$ & $c_2$ \\ 
\midrule
1 & 46 & 0.12 & 0.65 & 0.15 & 0.06 & 0.68 & 1.55 & 1.45 \\
2 & 50 & 0.1 & 0.7 & 0.1 & 0.1 & 0.729 & 1.494 & 1.494 \\
3 & 40 & 0.1 & 0.7 & 0.1 & 0.05 & 0.7 & 1.5 & 1.5 \\
4 & 50 & 0.15 & 0.6 & 0.3 & 0.1 & 0.6 & 1.8 & 1.8 \\
5 & 40 & 0.15 & 0.6 & 0.2 & 0.05 & 0.65 & 1.6 & 1.4 \\
6 & 40 & 0.12 & 0.63 & 0.2 & 0.02 & 0.68 & 1.7 & 1.3 \\
\bottomrule
\end{tabular}
\end{table}

%\begin{figure*}[t]
%\centering
%\subfloat[]{\includegraphics[width=0.3\textwidth]{FLeps2}} \hfil
%\subfloat[]{\includegraphics[width=0.325\textwidth]{FLeps3}}
%\captionsetup{justification=raggedright, singlelinecheck=off}
%\caption{(a) High-altitude view from front; (b) High-altitude view from back.}
%\label{trajectory} % 文中引用该图片代号
		%\vspace{-0.5cm}
%\end{figure*}

\begin{figure}[t]
    \centering
    \includegraphics[width=0.7\linewidth]{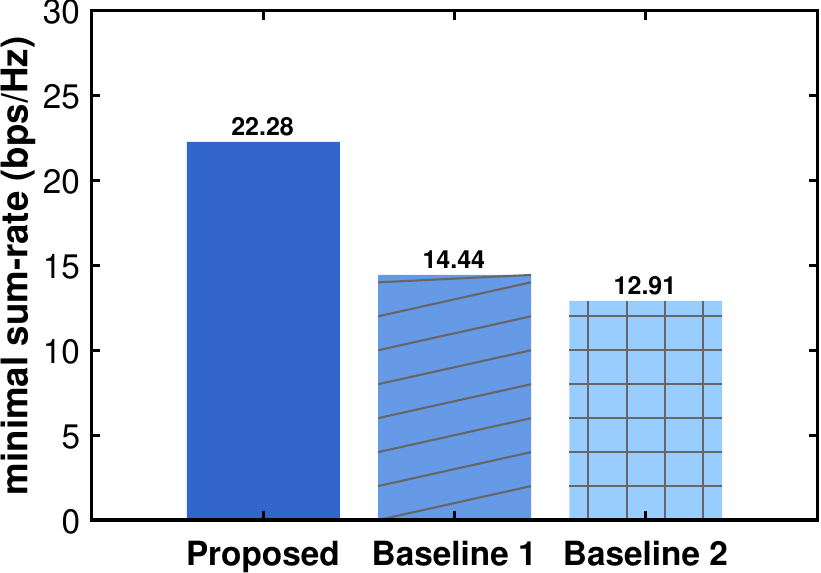}
    \caption{Performance comparison against baselines.}
    \label{fig2}
\end{figure}
\begin{figure}[t]
    \centering
    %\begin{minipage}[b]{0.48\linewidth}
      %  \includegraphics[width=\linewidth]{FLeps2}
      %  \caption{High-altitude view.}
      %  \label{High-altitudeview}
    %\end{minipage}
   % \hfill
    %\%begin{minipage}[b]{0.485\linewidth}
        \includegraphics[width=0.9\linewidth]{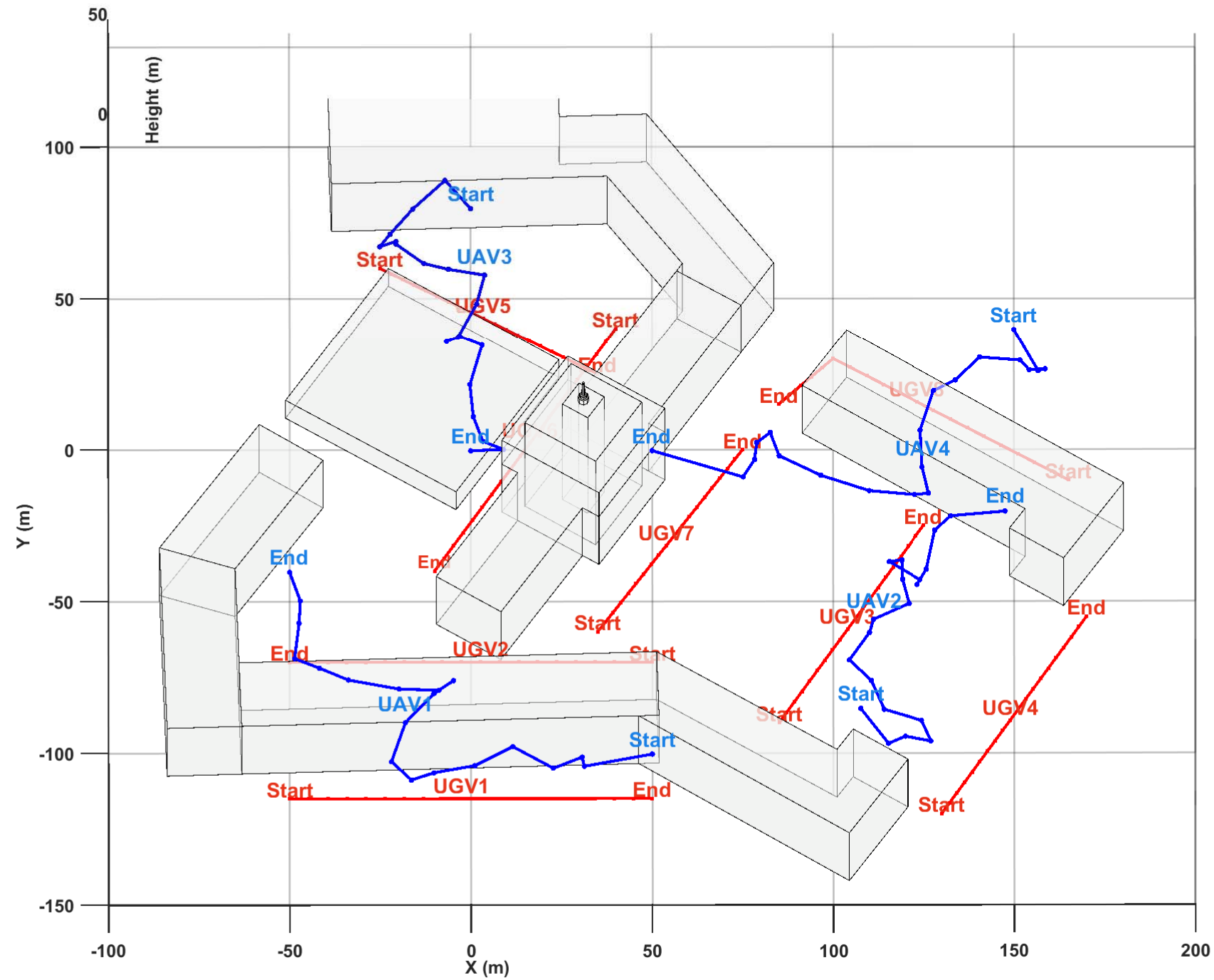}
        \caption{Visualization of the WS-PSO-CM optimized UAV 3D trajectories by proposed hyper-parameters.}
        \label{Topview}
    %\end{minipage}
\end{figure}

\section{Conclusion}
The proposed LLM agent for hyper-parameter optimization in the WS-PSO-CM algorithm significantly enhanced UAV trajectory and communication optimization, achieving performance gains of $54.33\%$ and $72.61\%$ over heuristic and random approaches, respectively. The success of this framework highlighted its broader applicability to related problems, such as multi-UAV coordination, dynamic resource allocation, radio map construction, and reinforcement learning tasks, where automated and intelligent tuning can replace manual intervention. By integrating LLM capabilities with optimization algorithms, LLM agents offered a scalable and efficient solution for complex systems, paving the way for advanced automation in wireless communication, autonomous systems, and beyond. 

\bibliographystyle{ieeetr} 
\bibliography{ref}

 \end{document}